\documentclass[twocolumn,prl,preprintnumbers]{revtex4}

\usepackage{times}
\usepackage{graphicx}
\usepackage{dcolumn}
\usepackage{bm}

\begin{document}

\title{\huge Probing complex RNA structures by mechanical force}

\author{S.~Harlepp, T.~Marchal, J.~Robert\footnote{Corresponding
  author: robert@fresnel.u-strasbg.fr}, J-F.~L\'eger, A.~Xayaphoummine,
  H.~Isambert\footnote{Corresponding author: herve.isambert@curie.fr
    ~~~On leave for Institut Curie, Section de Recherche, 11 rue P. \&
    M. Curie, 75005 Paris, France.} and D.~Chatenay} 

\vskip 0.2cm
 
\affiliation{Laboratoire de Dynamique des Fluides Complexes, CNRS-ULP,
  Institut de Physique, 3 rue de l'Universit\'e, 67000 Strasbourg, France} 


\maketitle

\noindent
{\bf
RNA secondary structures of increasing complexity are probed 
combining single molecule stretching experiments 
and stochastic unfolding/refolding simulations.  
We find that force-induced unfolding pathways cannot 
usually be interpretated by solely  invoking successive 
openings of native helices. Indeed, typical force-extension 
responses of complex RNA molecules are largely shaped by stretching-induced,
long-lived  intermediates including non-native helices. 
This is first shown for a set of generic structural motifs found in 
larger RNA structures, and then for  {\em Escherichia  
coli}'s 1540-base long 16S ribosomal RNA, which exhibits a surprisingly  
well-structured and reproducible unfolding pathway under mechanical 
stretching. Using out-of-equilibrium stochastic simulations, we demonstrate
that these  experimental results reflect the slow relaxation of RNA structural
rearrangements. Hence, micromanipulations of single RNA molecules probe both 
their native structures and long-lived intermediates, so-called ``kinetic
traps'', thereby capturing --at the single molecular level-- the hallmark of
RNA folding/unfolding dynamics.}  

\small

\vspace{.2 cm}



\vspace{0.6cm}
\noindent
{\large \bf Introduction}

Recent developments of micromechanical experiments on single 
biomolecules  
have provided structural insights into alternative 
structures of DNA\cite{chatenay,bustamante,leger1,allemand} and 
mechanical properties of proteins\cite{prot1,prot2,prot4}.
In principle, such techniques could also provide new tools 
to probe RNA structures which remain 
by and large refractory to many crystallization schemes. 
However, this prospect requires one to relate mechanically induced 
unfolding pathways to RNA structural features.
Although it could be done successfully for small RNA structures
by solely invoking successive openings of native 
helices\cite{liphardt1},
probing more complex RNA structures by mechanical force is expected 
to involve non-native structural rearrangements of the initial 
secondary structure upon stretching\cite{montanari,bundschuh,lubensky}.
Local rearrangements, such as the formation of simple stem-loops,
occur quite fast ($<$ 1 ms) under low pulling force (or in the absence 
of force) and the number of possible hairpins (with small loop) is 
proportional to the
length of stretched region of the RNA molecule.  
Thus, alternative hairpins, not present on the initial structure, should
inevitably form under partial stretching of long RNA molecules 
({\it e.g.}, $>$ 1000 nucleotides). Conversely, more global rearrangements, 
which involve
the coordonated removal and formation of different sets of helices, might 
occur much more slowly ({\it e.g.}, after a few 
minutes)\cite{williamson1,woodson}. 
Hence, under typical pulling rates ({\it i.e.}, full extension within a 
few seconds), most stretching experiments likely occur under
out-of-equilibrium conditions and should exhibit unfolding/refolding
hysteresis curves.

To study the full 
potential and limitations of these micromechanical techniques 
so as to probe complex RNA structures, we have combined single RNA molecule
stretching experiments  and out-of-equilibrium stochastic simulations.
Three small artificial structures, 
{\sf \small M1}, {\sf \small M2} and {\sf \small M3} (Fig~1),
representing prototypes for the main structural modules of larger
RNA secondary structures, were first designed and studied in details. 
The mechanical response
of {\em E.~coli}'s 1540-base long 16S ribosomal RNA was then  studied
using the same experimental setup and a somewhat simplified numerical
approach. 
The generic structural motifs {\sf \small M1}, {\sf \small M2} and {\sf \small M3}, correspond to three 
different arrangements  of two 15 base pair long helices 
consisting
almost exclusively of either GC or AU base pairs, Fig~1.
{\sf \small M1} corresponds to 
two adjacent stem-loops
with respect 
to the external single strand joining the molecule ends. 
By contrast, {\sf \small M2} 
and {\sf \small M3} present the same nested organisation with 
either the strong (GC) helix or the weaker (AU) helix connected to the
external single strand. 
\begin{figure}
\includegraphics{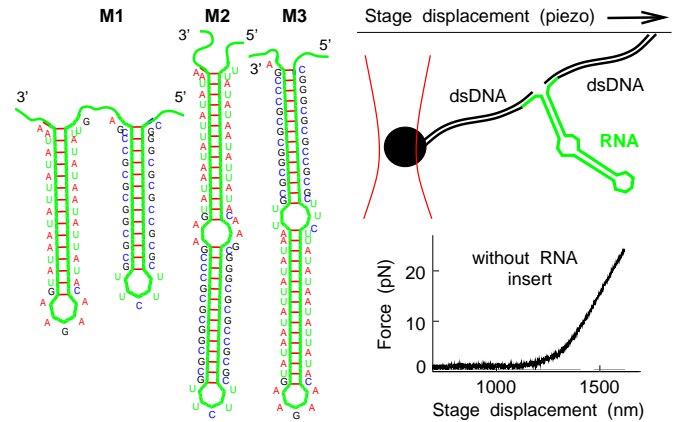}
\caption{\label{fig:wtgI} 
\small The three structural motifs with the schematic setup (see Materials
  and Methods) and a force-extension curve of two ligated pUC19 in the 
absence of RNA insert.}
\vspace{-.3cm}
\end{figure}

The 5' and 3' ends of either these small RNA motifs or {\em E.~coli}'s 16S
rRNA were hybridized  
to two pUC19 dsDNA extensions labelled, respectively, with biotin 
and digoxygenin (see Materials and Methods).
The force-extension experiments were then done by grafting the ends of these 
extended molecular constructs between the antidigoxygenin coated glass 
surface of a capillary and a micrometer size silica bead coated with  
strepavidin. The capillary was moved by a piezo-electric stage (50 to 300~nm/s)
 and the 
resulting force exerted by the molecule on to the bead was measured with
an optical tweezer.

\begin{figure}
\includegraphics{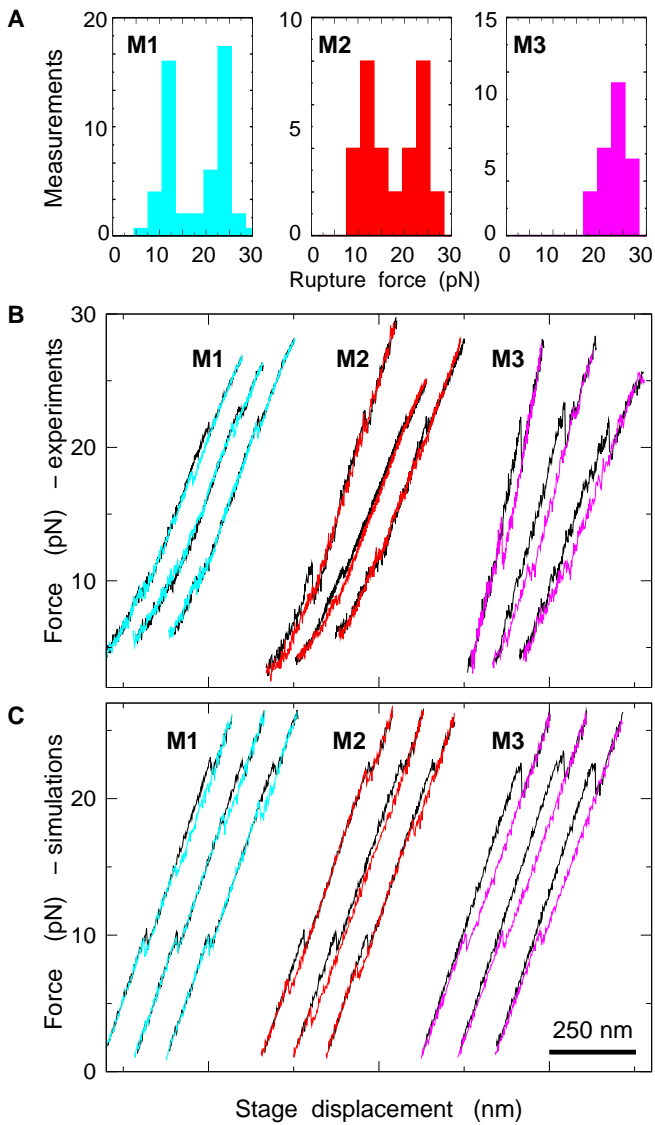}
\caption{\label{fig:wtgI} 
{\bf A} {\small Histogram of the measured rupture forces for 
the three structural 
motifs  (3pN bins).} 
{\bf B} {\small Experimental force-extension curves of the three structural 
motifs. Note hysteresis between unfolding (black) and refolding curves 
(color).} {\bf C} {\small Corresponding
stochastic simulations. The mechanical stiffness of the optical tweezer
and the wormlike chain elasticity of the pUC19 dsDNA 
extensions (which curves the experimental force-extension slope at low 
stretching force) are combined for simplicity
into an effective stiffness with a slop fitted on the experimental curves 
(0.1~pN/nm).}}
\vspace{-.3cm}
\end{figure}
\begin{figure}
\includegraphics{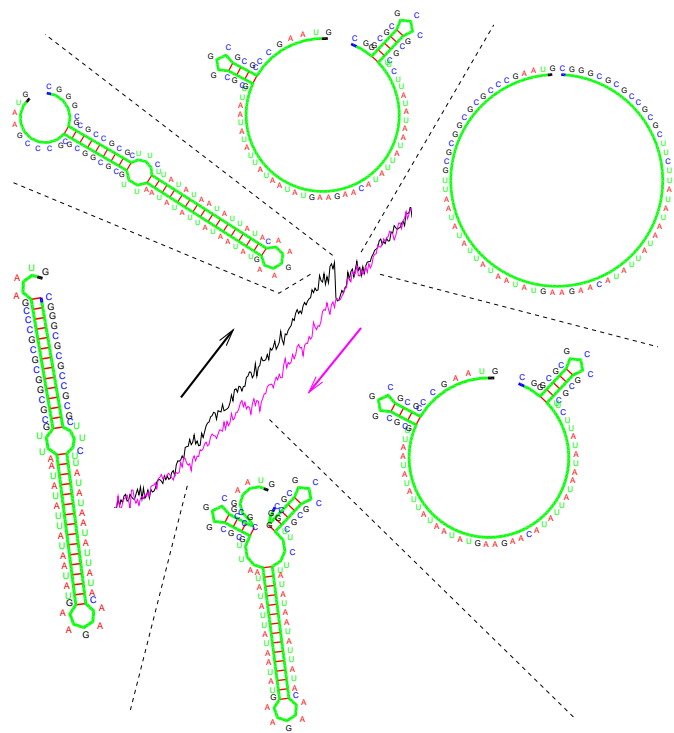}
\caption{\label{fig:wtgI} 
\small Interpretation of the experimental unfolding/refolding hysteresis for 
the structural motif M3 (see text). Regions under tension are drawn
on a circle for convenience\protect\cite{evers}.}
\vspace{-.3cm}
\end{figure}

\vspace{0.6cm}
\noindent
{\large \bf Results}

\vspace{0.1cm}
\noindent
{\bf Single molecule stretching experiments of small RNA motifs}

When structural motifs {\sf \small M1, M2} or {\sf \small M3} are inserted
in the molecular construct,  one or two force drops occur on the
force-extension curve, Fig~2. 
A histogram of the rupture force and a
set of unfolding and refolding force-extension curves 
are shown for each motif on Fig~2A and Fig~2B. 
For each set, variations between force-extension curves correspond to 
stochastic fluctuations between either successive stretchings on the same 
molecule or different experiments on equivalent molecules.

A comparison of the different rupture force histograms and the corresponding 
unfolding curves (black on Fig~2B) shows that 
{\sf \small M1} and {\sf \small M2} present very similar unfolding responses
with  two sequential drops or inflexion regions around 11~$\pm$~3~pN and 
22~$\pm$~3~pN,  whereas {\sf \small M3} presents a single and larger 
force drop at about 22~$\pm$~3~pN.
These values are in very good agreement with ref.\cite{gaub} although these
latter  experiments concern the opening of DNA hairpins.
For {\sf \small M1} and {\sf \small M2}, these results can be simply
attributed to the first  
opening of the weak (AU) helix followed by the stronger (GC) helix at a 
higher applied force. Indeed, the applied tension being uniformly distributed
along the external single strand joining the molecules ends, the weaker (AU)
helix is expected to break first on {\sf \small M2}, while it should certainly
do  so by construction on {\sf \small M1}.  
Besides, by calibrating the stiffness of the optical trap, both force drops 
on these
curves can be converted into a distance released by the molecule, taking into
account the angular inclination of the setup (30$^\circ$ to 40$^\circ$). 
This corresponds to the expected 20~nm in both cases. 
Substracting the net
free-energy contribution stored in the stretched single strand\cite{cocco}, 
we find in term of pairing energy, around 1.7~kT/bp for AU and 3~kT/bp for GC,
in good agreement with known parameters\cite{turner}.
By contrast for {\sf \small M3}, the strong (GC) helix shields the weaker (AU)
stem 
from the applied force and no significant unzipping is observed until the whole
molecule suddenly unfolds at the critical force to break GC stacking base 
pairs. 


The refolding curves for {\sf \small M1} (blue) and {\sf \small M2} (red) show
most  
often a small hysteresis below the force drop associated with the 
strong (GC) stem's opening. For {\sf \small M2}, a second small refolding 
hysteresis occurs also usually below the 
force drop associated with the weaker (AU) stem's opening.  
By contrast, a much stronger hysteresis is systematically observed for 
{\sf \small M3} (magenta), even at the 
lowest loading rate achieved, 3~pN/s. Moreover, in this case,  
the refolding
event around 10 pN does not usually fold back onto the initial stretching 
curve. This suggests that the stretching of {\sf \small M3} involves
long-lived intermediate structures including non-native helices (see Fig~3 
and next section on stochastic simulations of small RNA motifs).
Still, all three molecules eventually fold back in their initial native 
structure after a few seconds, as shown by the reproducibility of force
extension curves in successive pulls on the same molecule.

\vspace{0.6cm}
\noindent
{\bf Stochastic unfolding/refolding simulations of small RNA motifs}

We have performed stochastic simulations of these out-of-equilibrium 
unfolding/refolding experiments for the short {\sf \small M1, M2} and 
{\sf \small M3} structural motifs. 
The heart of the numerical method, following the approach detailed 
in\cite{isambert}, consists in simulating the stochastic unfolding and 
refolding of helices not only present on the initial RNA structure but also 
for {\it all other helices} which can possibly pair on 
the RNA sequence of interest (see also RNA {\it Kinefold} server at 
{\sf \small http://kinefold.u-strasbg.fr }). 
Common pseudoknots (i.e., helices interior to loops) are also allowed
following the structural modeling approach proposed in\cite{isambert}.
In addition, the 
region of the RNA structure under {\it direct} mechanical tension 
(corresponding to the ``on-net'' backbone in\cite{isambert}) 
is modelled as an inextensible wormlike chain
with a 1.5 nm persistence length\cite{bustamante} and 0.7 nm/base contour
length\cite{contour}. Stretching is induced by 
a slowly varying rigid constraint on the
end-to-end distance of the RNA-dsDNA-tweezer construct 
(rate $\pm$~300~nm/s in $\pm$~2~nm steps).
It is also important to take into account the acquisition rate (300 Hz) and 
to  model the statics and dynamics of the optical 
tweezer trap, although time scale separation allows to consider that the
micromechanical setup responds to a slow time average of the fast 
RNA dynamics, which corresponds to stochastic closing and opening
of single helices\cite{isambert}. 
To avoid overfitting with non essential parameters, we have simply modelled
the trapped bead and the two dsDNA extensions of the construct as an 
ideal spring with a slow viscous relaxation time (1 ms)
\noindent  and an effective 
stiffness (typically 0.1 pN/nm) fitted on the individual force extension 
curves.

The simulated force-extension responses for the {\sf \small M1, M2} 
and {\sf \small M3}  
motifs (Fig~2C) are in good agreement with the unfolding and refolding 
experimental results (Fig~2B). In particular, they allow for the
identification 
of likely intermediate structures involved in the refolding hysteresis,
which primarily correspond to the formation of two non-native 
helices originating from each 
strand of the strong 
(GC) helix, Fig~3. The transition
from these alternative helices back to the strong initial (GC) stem is 
facilitated  
under high external force, hence the small hysteresis for
{\sf \small M1} and {\sf \small M2}. By contrast for {\sf \small M3}, the transition can only 
occur at a lower force after the weak (AU) stem has refolded and is, 
therefore, slower, as observed experimentally.
Note,  the small experimental differences in the hysteresis 
responses of {\sf \small M1} and {\sf \small M2} are well reproduced on 
their simulated force-extension 
curves, suggesting that elementary unfolding/refolding events are
reliably captured by these stochastic simulations devised to probe RNA 
molecular dynamics on second to minute time scales. It should be emphasized
that such long time scale simulations could not be achieved for other
molecules, such as proteins, for which elementary unfolding/refolding 
transitions are not so easily defined and also much 
more frequent. For instance, the best molecular dynamic simulations of 
proteins are currently limited to around 100~ns\cite{fersht}.

The force-induced unfolding of these three generic structural motifs 
{\sf \small M1, M2} and {\sf \small M3},
reveals the potential and limitations of  single molecule experiments 
to probe the main folded features of more complex RNA structures.
The comparison between {\sf \small M2} and {\sf \small M3}'s force-extension
responses illustrates that 
the order of helix stability along a single secondary structure branch can be 
readily identified,  
while the bifurcation arrangements of helices or the  presence of  
multibranched loops are not so easily distinguished from single 
branches with increasing helix stability (as in {\sf \small M1} versus 
{\sf \small M2}).  Moreover, the formation of non-native rearrangements 
under stretching 
likely affects the force-extension responses of most RNA structures 
(as for {\sf \small M3}).

In this context, combining experimental and numerical approaches 
to study RNA mechanical unfolding pathways seems promising insofar
as transient structural rearrangements (under stretching) are difficult
to probe with traditional chemical or enzymatic techniques. 

On the other hand, for long RNA molecules (e.g., $>$1000 bases),
it has been argued\cite{montanari,bundschuh,lubensky} that such
structural rearrangements under stretching should ultimately smooth out 
the observed characteristics 
completely by continuous adjustments to the applied constraint, assuming
that quasi-equilibrium stretching is achieved.

To investigate this issue and test whether large structures of biologically 
relevant RNA 
molecules are also amenable to convergent studies in both single 
molecule experiments and stochastic simulations, we decided to study the 
mechanical unfolding of {\it Escherichia coli}'s 1540-base 
long 16S ribosomal RNA.

\vspace{0.1cm}
\noindent
{\bf Single molecule stretching experiments of {\em E. coli} 16S rRNA}

\begin{figure}
\includegraphics{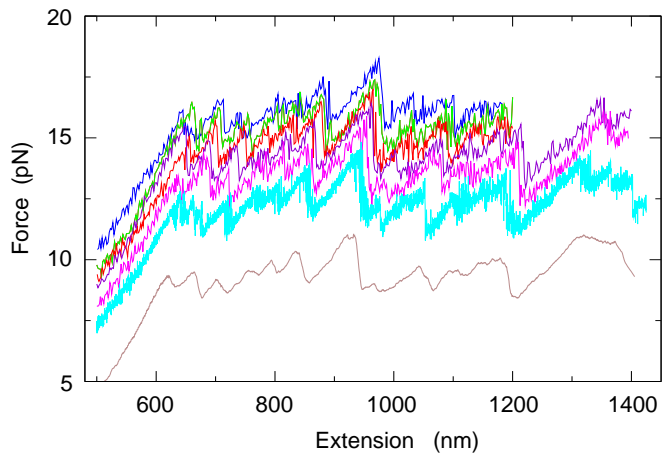}
\caption{\label{fig:wtgI} 
\small 
Experimental unfolding of {\it Escherichia coli}'s 1540-base long 16S
ribosomal RNA by mechanical stretching (rate 300nm/s).
Colors correspond to successive stretching rounds of the {\it same} 
molecule  (refolding hysteresis are not shown for
clarity). 
An increasing maximum extension was applied at successive stretching/refolding
rounds to avoid early breakage by overstretching.
As a result, the RNA molecule was not entirely unfolded until the sixth 
stretching/refolding round.  
Force-extension curves are slightly shifted vertically and
horizontally to best display the overall reproducibility between successive
extensions. 
The mechanical unfolding over the full extension range of the molecule
presents a characteristic unfolding plateau between 11 and 15pN.
This is the mechanical unfolding  signature of {\it E. coli}'s 16S rRNA.  
The brown curve (bottom) corresponds to the average
of the colored curves above.
}
\end{figure}
\begin{figure}
\includegraphics{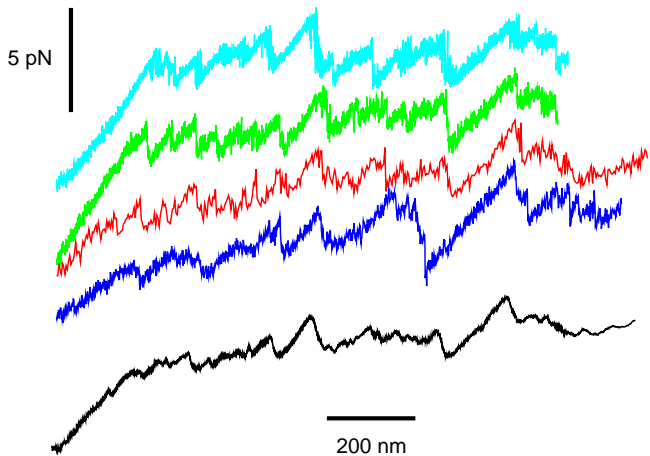}
\caption{\label{fig:wtgI} 
\small 
Reproducibility of the experimental unfolding of {\it E. coli} 16S
ribosomal RNA under mechanical stretching (rate 50-300nm/s).
Colors correspond to stretching responses of {\it different}
16S molecules taken from {\it independent} sample preparations and 
{\it independent} micromechanical experiments 
(refolding hysteresis are not shown for clarity). 
The force-extension curves have been shifted vertically and
horizontally to best display the overall reproducibility between these
independent measurements. The black curve (bottom) corresponds to the average
of the four colored curves above. 
}
\vspace{-.3cm}
\end{figure}

\begin{figure}
\includegraphics{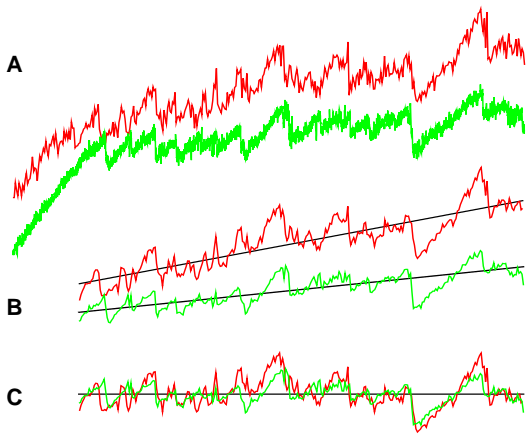}
\caption{\label{fig:wtgI} 
\small 
Statistical comparison of stretching responses.
{\bf A}: Two {\it independent} unfolding curves of 16S rRNA with different
sampling rates (green and red curves from Fig~5).
{\bf B}: Extraction of unfolding plateau signals from the overall non-specific
stretching curves and uniform smoothing (N=300 regularly sampled points).
This enhances correlation 
sensitivity 
to  the specific unfolding signatures  
around the median line fits ({\it i.e.,} line minimizing
absolute deviations\protect\cite{recipes}).
{\bf C}: Deviations from the median line fits are used to calcule the 
{\it relative} Spearman correlation coefficient
$r_s$ \protect\cite{recipes} between  unfolding curves  (see text).
In the example shown, $r_s$=60.5\% which
corresponds to a very good correlation between the two experimental deviations 
relative to the median line fits.
Evaluating, instead,  {\it absolute}  
correlations between 
actual unfolding responses  (above curves) yields an even larger, 
{\it yet less discriminating}, Spearman coefficient $r_s$=90\%, due to the 
small positive slopes of both unfolding plateaux. 
}
\vspace{-.3cm}
\end{figure}

The force-induced stretching of 
{\it E. coli} 16S ribosomal RNA was studied 
using a similar molecular construct and micromechanical setup as 
for the stretching of the small structural modules {\sf \small M1, M2} and {\sf \small M3} 
(see Materials and Methods). 
No ribosomal proteins which associate to 16S rRNA to form the 30S 
subunit\cite{ramakrishnan} of the ribosome\cite{noller1}  were included
for these stretching experiments.  
 As the piezo stage is displaced, the force begins to rise due to the elastic
 response of the DNA handles.  The results on Figs~4-5
 show a well-structured and reproducible unfolding 
pathway under mechanical stretching, 
in about 50\% of the tested constructs for which more than two 
unfolding/refolding rounds could be performed before molecular breakage. 
In these cases (total 44 stretching curves), a $\sim$~1~$\mu$m-long quasi
 plateau is observed around 11-15 pN, with force fluctuation amplitude of 
about 20\%. This signal is the signature of 16S rRNA unfolding by mechanical
force. Other stretching curves exhibit somewhat more erratic
behaviors, presumably due to non-specific interactions 
of the construct with the glass surface of the capillary (data not shown)
Extension beyond the unfolding plateau corresponds to the combined elastic
response of the dsDNA handles and the opened ssRNA molecule. 
Most refolding curves exhibit strong hysteresis depending on stage velocity 
(50-300 nm/s).

We have quantitatively evaluated the statistical reproducibility of unfolding
curves between successive stretchings of the {\it same} 16S molecule (Fig~4) 
and between {\it independent} unfolding curves from {\it different} 16S 
molecules (Fig~5) ({\it i.e.},  {\it different} sample preparations and 
{\it different} micromechanical experiments). 
The analysis is based on 
the calculation of Spearman  nonparametric correlation coefficient $r_s$ 
\cite{recipes}:
$r_s=\Sigma_i^N(R_i-\bar R)(S_i-\bar S)/(\Sigma_i^N(R_i-\bar R)^2
\Sigma_i^N(S_i-\bar S)^2)^{1/2}$, where $R_i$, $\bar R$ and $S_i$, $\bar S$
are chosen as the rank-ordered {\it deviations} and averages from the {\it
  median line fits} of the unfolding plateaux (Fig~6).  
Such {\it relative} Spearman correlation coefficient
is much more sensitive to the specific unfolding signals,
as compared to the {\it absolute} Spearman correlation
of the actual unfolding curves which yields higher, yet less
discriminating correlation coefficients (see Fig~6 for details).

In the context of comparing 16S unfolding curves, we found that this relative
Spearman correlation coefficients 
correspond to good correlations above 50\% and 
excellent ones above 70\%, while $\vert r_s \vert<$15\% reflects little 
or no correlation between unfolding pathways irrespective of the
overall inclination of their unfolding plateaux. 
Stochastic reproducibility between successive stretchings of the {\it same} 
molecule is remarkably high ($r_s$=75$\pm$4.6\%, Fig~4) and still quite good
between  unfolding curves of {\it different} 16S RNA  ($r_s$=53$\pm$9.1\%,
Fig~5) despites  
inherent variations between {\it different} sample preparations and 
{\it different} micromechanical experiments. For instance, correlation between 
{\it independent} green and red unfolding curves on Fig~6 is: $r_s$=60.5\%,
while the correlation distributions of all {\it independent} curves of Fig~5 
(colors) with their {\it average} unfolding response (black)
is even higher: $r_s$=70$\pm$4.7\%.
See Table~1 for further correlation data and quantitative comparison with 
stochastic simulations.

Before discussing the stochastic unfolding/refolding simulations of 16S rRNA,
we want to emphasize that force fluctuations from the plateau median line 
cannot be 
attributed to dehybridization of the DNA handles for the following reasons: 
{\it i}) we never recorded such signals on simple pUC19 dimer without RNA 
insert (see Fig~1 inset). 
{\it ii}) experiments done by other groups with the same infrared laser power 
have preserved nucleic acids' integrity\cite{bustamante,svoboda}.
{\it iii}) recalling that we pull on opposite DNA strands, the average force 
magnitude at which the plateau appears is too low to originate from DNA 
denaturation\cite{chatenay,bustamante,leger1}. Moreover, no torque 
is applied on the molecules using our optical tweezer\cite{allemand}.
In addition, experimental
force fluctuations cannot correspond to the unzipping of a long structureless 
double stranded RNA molecules; an analysis based on G+C contents as
in ref.\cite{bock} does not account for the experimental signal, nor does a 
thermal equilibrium energy calculations. 

\begin{figure}
\includegraphics{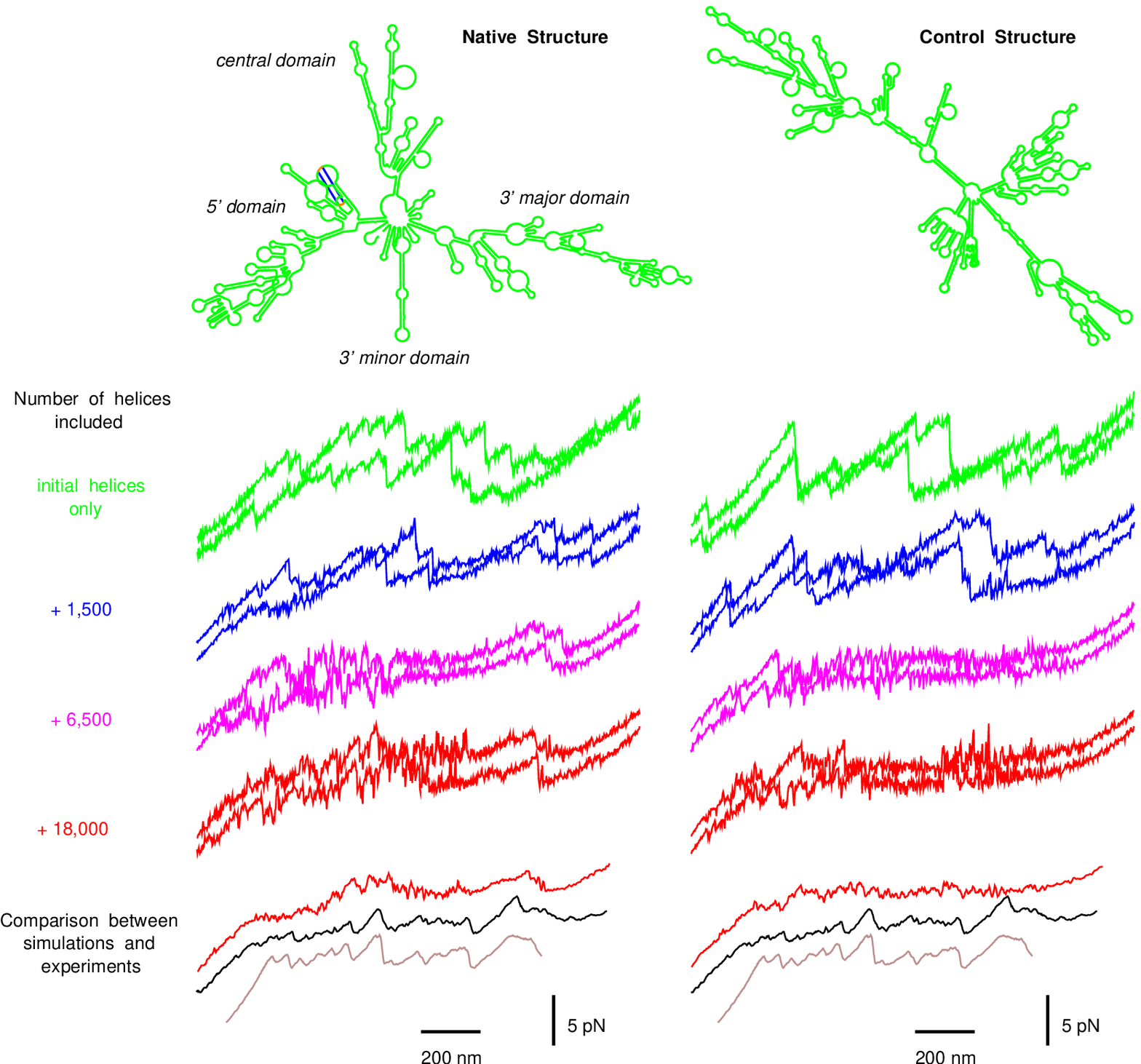}
\caption{\label{fig:wtgI} 
\mbox{
\small Simulated force-extension responses of {\it E.~coli}'s 
16S rRNA {starting from the known 
native structure}\protect\cite{gutell,ramakrishnan} (left) and from a low
energy}  
\mbox{ 
control structure (right).
Two stretching curves are plotted for each simulation conditions (various 
colors) to illustrate reproducibility.}
\mbox{ 
Cross-correlations between unfolding curves in each simulation
conditions and quantitative comparison with experiments are presented
in Table 1\\}
\mbox{ 
Red curves (see below) for the native structure (left) resemble most closely 
the experimental curves (brown from Fig~4 and black from Fig~5)\\}
\mbox{
also plotted for 
comparison.}\\
\mbox{
({\bf Green}):~the stochastic simulation is restricted to the sole helices 
formed on the initial structure assuming, in
addition, that those cannot refold \\}
\mbox{
once broken (78 helices for the native structure; 86 helices for 
the control structure).\\}
\mbox{
({\bf Blue}):~the initial helices and some 1,500 additional stems longer than
3bp and containing the most stable stack (5'-GC/\-GC-3') 
can form and}
\mbox{
break stochastically during stretching.\\}
\mbox{
({\bf Magenta}):~all additional helices longer than 3bp and stronger than 15kT
are also included; total: 6,500 helices.\\}
\mbox{
({\bf Red}):~all additional helices longer than 3bp and stronger than 10kT are
also considered; total: 18,000 helices. The lowest red curve corresponds}
\mbox{
to the average of four independent stretching simulations starting
  either from the native structure (left) or the control structure (right).} } 
\vspace{-.3cm}
\end{figure}

\vspace{0.6cm}
\noindent
{\bf Stochastic unfolding/refolding simulations of 16S ribosomal RNA}

The unrestricted stochastic simulations discussed above to model the 
mechanical unfolding and refolding of small structural motifs are
numerically unpractical in the case of much larger RNA structures like
those of ribosomal RNAs. 
Hence,  we have made the following three additional assumptions 
 to study the force-induced  stretching  of {\it E.~coli}'s 
1540-base long 16S ribosomal RNA:
{\it i)} the {\it initial} structure before mechanical stretching is
assumed known from independent sources; 
{\it ii)} Unfolding and
(re)folding dynamics is restricted to the formed helices under {\it direct} 
mechanical tension {\it and} to all potential helices that would be
under {\it direct} tension once formed. Hence, large scale structural 
rearrangements
can only originate and propagate from helices directly 
coupled to the
applied mechanical tension, as expected under strong
 stretching conditions; 
{\it iii)} For each intermediate structure along the
 unfolding
pathway, the actual base pair extent of
each helix under direct tension is {\it not globally optimized} to best
fit the end-to-end molecular extension imposed by the mechanical 
setup (this would become exponentially difficult in the number of such 
helices). Instead, a local heuristics extending the most stable 
base pair stacks and shrinking the weakest helix ends is used iteratively
to minimize free energy. 
This approach, which yields a linear optimization in the number of 
helices under direct
 tension, is usually very good as long as there are
few mutually incompatible helices competing for the same bases, a 
typical situation under strong mechanical stretching.
Overall, we found that these restricted stochastic simulations give virtually 
identical results for the small {\sf \small M1, M2} and {\sf \small M3} motifs (results not 
shown).

 \begin{table*}
\begin{ruledtabular}
 \caption{Statistical correlations between 16S 
rRNA mechanical responses under stretching. Spearman correlation 
coefficients $r_s$\protect\cite{recipes} are given between the deviations 
from the median line fits of the unfolding plateaux (see text and Fig~6).
Statistics are made from: 4 to 10 unfolding curves for each  
of the 4 stochastic simulation conditions ({\bf A}, see text and Fig~7);
{\it independent} experiments on 4 {\it different} 16S rRNA molecules 
({\bf B}, from Fig~5); and 6 successive stretchings from the {\it same} 
16S molecule ({\bf C}, from Fig~4). Curves are compared 
both between each other within each set (to evaluate ``stochastic 
reproducibility'') and with the experiment average response (black curve on
Fig~7).  \label{}} 
 \setlength{\extrarowheight}{4pt}
 \begin{tabular}{|c|c|c|c|c|} \hline
 {\bf 16S rRNA unfolding response}
 & \multicolumn{2}{c|}{\bf Stochastic reproducibility} &
 \multicolumn{2}{c|}{{\bf Comparison with experiment average} ({\it black})} \\
 Deviation from median line
 & \multicolumn{2}{c|}{Ensemble  cross-correlation (mean $\pm$ std dev.)} &  \multicolumn{2}{c|}{Correlation distribution  (mean $\pm$ std dev.)} \\
 \hline \hline
 {\bf ~A ~~~~~~~~~~Simulations}  ~including:~~~~~~~~~~~~~~
& ~~~~Native structure~~~~~~ 
&  Control structure 
& ~~~~~Native structure~~~~~~ 
& Control structure  \\  \hline
 ~~$\sim$ 80 initial helices \hfill({\it green})~~
& 20  ~$\pm$~ 21~~\% 
& 20  ~$\pm$~ 20~~\%  
& 23 ~$\pm$~ 32~~\% 
& 12 ~$\pm$~ 12~~\% \\  \hline
 ~~1,500 helices {\it incl.} 5'GC/GC3'\hfill({\it blue})~~
& 22  ~$\pm$~ 28~~\% 
& 48  ~$\pm$~ 21~~\%  
& 31 ~$\pm$~ 30~~\% 
& -1.7 ~$\pm$~ 8.5~~\%  \\  \hline
 ~~6,500 helices $>$15kT, 3bp \hfill~({\it magenta})~~
& 12  ~$\pm$~ 22~~\% 
& 27  ~$\pm$~ 12~~\%  
& 21 ~$\pm$~ 17~~\%
& 14 ~$\pm$~ 8.6~~\%  \\  \hline
 ~~18,000 helices $>$10kT, 3bp\hfill({\it red})~~ 
& {\bf 57 ~$\pm$~ 3.8}~~\% 
& 15 ~$\pm$~ 10~~\%  
& {\bf 47 ~$\pm$~ 8.8~~\%}
& {\bf -7.2 ~$\pm$~ 12~~\%}  \\ 
 ~~simulation average \hfill({\it red})~~
& {\it NA}   
& {\it NA} 
& {\bf 55~~\%}
& {\bf -1.1~~\%} \\  
 \hline \hline \hline
 {\bf B~~ Experiments} on {\it different} molecules~~ & \multicolumn{2}{c|}{{\bf 53 ~$\pm$~ 9.1~~\%}} &  \multicolumn{2}{c|}{{\bf 70 ~$\pm$~ 4.7~~\%}} \\
 ~~experiment average \hfill({\it black})~~ & \multicolumn{2}{c|}{\it NA} &
 \multicolumn{2}{c|}{100~\% ~~(reference)} \\
\hline
 {\bf C~~ Experiments} on the {\it same} molecule~~~& \multicolumn{2}{c|}{{\bf 75 ~$\pm$~ 4.6~~\%}} &  \multicolumn{2}{c|}{{\bf 71 ~$\pm$~ 4.7~~\%}} \\
 ~~experiment average \hfill({\it brown})~~ & \multicolumn{2}{c|}{\it NA} &
 \multicolumn{2}{c|}{\bf 75~~\%} \\\hline 
 \end{tabular}
 \end{ruledtabular}
 \end{table*}

Adopting this heuristic numerical approach
for the bare {\it E.~coli}
\mbox{~~~~~~~~~~~~~~~~~~~~~~~~~~~~~~~~~~~~~~~~~~~~~~~~~~}

\vspace{20.4cm}
\noindent
 16S rRNA, we simulated the force-induced unfolding pathway
starting either from the known native secondary structure  
inside the ribosome\cite{ramakrishnan,noller1} or from 
a low free-energy structure predicted by mfold, referred at,  hereafter, 
as the ``control structure'' 
({\sf \small http://bioinfo.math.rpi.edu/$\sim$mfold/}).
The comparison between the force-extension responses of these two
structures was primarily intended to probe the stochastic simulation's
sensitivity to the initial structure.
In both cases, the role of helices not initially formed on
the starting structure was studied, by allowing a variable 
number of helices to form and break during different stochastic simulations.
The results in Fig~7 and Table~1 show that a reasonable agreement
exists between the experimental measurements (black and brown curves) 
and 
the simulated force-extension responses {starting from the known 
native structure}, when {\it all helices longer than 3bp and more stable
than 10kT} (i.e., 6 kcal/mol) are included {\it a priori} in
the simulations (red curves). This demonstrates that some of these 
18,000 different non-native helices play a significant structural role 
along the unfolding pathway. More quantitatively,
cross-correlations amonsgt 4 independent simulated stretching curves 
({\it i.e.}, about 3 weeks of CPU on a 1.2GHz PC) reveal
a good stochastic reproducibility in these simulations starting from the
native structure and including about 18,000 possible helices: 
$r_s$=57$\pm$ 3.8\%. This is comparable to
observed variations between  experimental unfolding curves (see Table~1). 
Then comparing these individual simulations with the 
experiment average curve (black curve on Fig~5 from 4 independent experiments),
we obtain a significative correlation coefficient: $r_s$=47$\pm$8.8\%, while
correlating the experiment average (black)  directly to the simulation average
clearly reflects common features between the experimental response and 
the simulated unfolding pathway curve starting from the native structure:
$r_s$=55\%. 
In addition, restricting simulations to the 6,500 possible
helices longer than 3bp and stronger 
than 15kT (magenta curves) or including even fewer helices (blue and green 
curves), produces marked differences on the simulated stretching 
curves (Table 1 shows lower averages and larger standard deviations for the 
simulation stochastic reproducibility and for the correlations with 
experimental response). 
By contrast, equivalent stochastic simulations starting from the control 
structure (Fig~7)  present 
clearly distinct results from experimental observations, 
{\it i.e.}, $\vert r_s \vert<$15\% (even for a large number of possible
helices included in the simulations).
The fact that the stochastic reproducibility of these control simulations 
happens to decrease with the number of possible helices taken into 
account (Table~1) reflects the concomitant decrease of specific unfolding
signal relative to the median line fit of the plateau (Fig~7). A decreasing 
signal over noise ratio naturally leads to a lower reproducibility
of the simulated curves. The same trend is also visible between blue and
magenta curves for simulations starting from the native structure.

Fig~8 compares more closely a simulated force-extension 
response of the known native structure (red) 
and an experimental stretching curve (black). 
Again, both simple visual comparison and calculation of their correlation 
coefficient as above (here $r_s$=61\%) strongly suggest that   
the experimentally probed structure  
shares, indeed, more structural features with the actual native structure 
than with 
the control structure (Fig~7), in spite of the absence of ribosomal proteins 
in 
these single molecule stretching experiments of {\it E. coli}'s 16S rRNA. 
Analysing the unfolding pathway during the simulated force-extension response 
reveals that
the main predicted unfolding events (corresponding to abrupt
force drops on the red curve) 
are either related to
the {\it cooperative opening of several 
native helices} (as in the unfolding of {\sf \small M3}) or to the {\it
simultaneous rearrangements of mainly non-native helices} leading to a
stepwise  increase of the predicted 
extension of the molecule along the direction of pulling (violet curve).
This is illustrated with 12 successive snapshots of intermediate 
structures along the stretching-induced unfolding pathway.
In particular, the 3' major ({\it III}) and 3' minor ({\it IV}) domains are 
shown to break and partially 
rearrange at the start of the stretching plateau (intermediates 2 to 4) 
while the 5' domain ({\it I}), partially unfolded between intermediates 
4 and 5, remains then largely intact until most other native
and non-native helices have been opened under stretching (intermediate 11).
Finally, the central domain ({\it II}) exhibits a more distributed unfolding 
fate which extends from intermediates 1 to 10. 
Hence,  mechanical breaking of the native structure does {\it not} 
occur  through successive openings of entire native domains. 
Instead, native helices contribute to a more complex (yet largely reproducible)
sequence of force drops, reflecting also the rearrangements of non-native
helices. For instance, this is the case for the recorded signal between 
intermediates 8 and 10 which is largely caused by successive rearrangements 
of weak non-native helices between 10kT and 15kT (compare magenta and red 
curves for the native structure on Fig~7 and  experimental and simulated
curves on Fig~8).
In retrospect and more generally, these results
underline the possible pitfalls in attempting to assign
specific structural features of large RNA molecules
by studying the mechanical unfolding of their independently folded
domains  separately.

\begin{figure}
\includegraphics{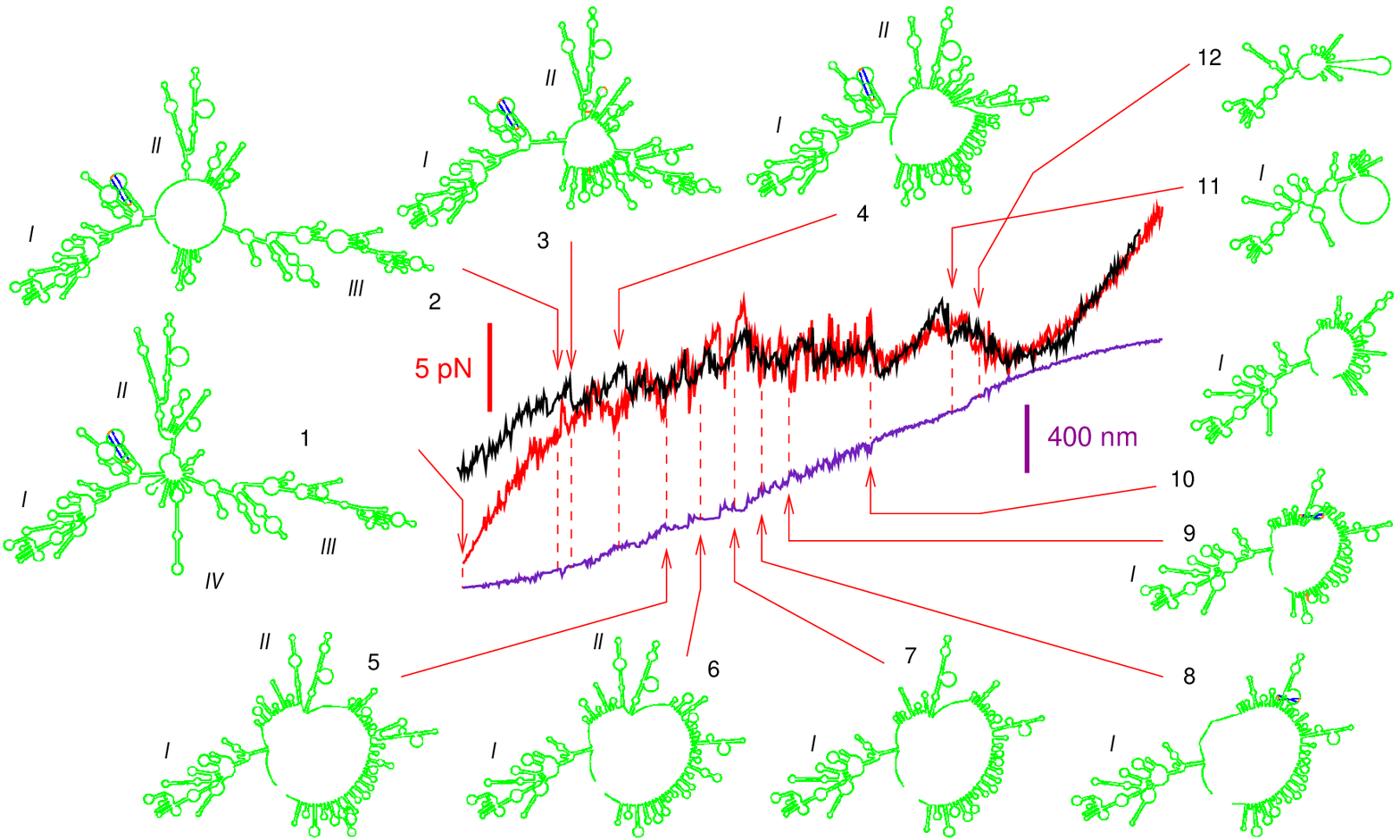}
\caption{\label{fig:wtgI} 
\noindent
\mbox{ 
{\small Comparison between simulated force-extension response from the known
native structure\protect\cite{gutell,ramakrishnan} (red) and an experimental stretching}} 
\mbox{{\small curve (black) of {\it E.~coli}'s 1540-base long 
16S ribosomal RNA. 
Spearmann correlation coefficient on this example: $r_s$=61\%. The simulated}} 
\mbox{{\small end-to-end molecular extension of the 16S rRNA is also plotted 
(violet). Twelve intermediates 
on the simulated unfolding pathway are drawn}}
\mbox{{starting with the known native structure. Single stranded regions under tension
    are not drawn for convenience, hence the overall decreasing size}}
\mbox{{of the structure under stretching.}} 
}
\vspace{-.3cm}
\end{figure}

\vspace{0.6cm}
\noindent
{\large \bf Discussion}

We have measured the force range to unfold  RNA secondary 
structures by mechanical stretching experiments. 
It extends from 10 pN for AU rich to 25 pN for GC rich regions 
in agreement with intermediate values reported for intermediate G+C contents.
We also showed that non-native rearrangements have  a large 
influence on force-extension measurements of complex RNA structures, as in
the case of {\it E. coli}'s 16S rRNA presented here. 

Interestingly, this force-induced unfolding process of the bare 
16S rRNA's domains seems to mirror, only in reverse order, the 
predominant 5' to 3' polarity of the {\it in vitro} assembly of 
16S rRNA into 30S ribosomal subunits\cite{noller}.   
From a more general perspective, the high reproducibility of 
the mechanical unfolding curves shown here ({\it e.g.}, Figs~4-5) sharply
contrasts with the multiple folding and misfolding pathways 
usually experienced by RNA molecules of this size during thermal 
renaturation.
This reflects the fact that unfolding/refolding pathways 
under mechanical constraint solely explore a restricted number of 
possible intermediate structures, as compared to unconstrained 
denaturation/renaturation folding experiments.
In other words, single molecule unfolding and refolding experiments 
under mechanical control probe  particular, well-defined pathways
due to the slowly varying external constraint applied onto the ends of 
the RNA molecule. In addition, we found that the overall unfolding curves
did not critically depend on the rate of pulling used 
(typically 300~nm/s); 
for instance, imposing 
\mbox{~~~~~~~~~~~~~~~~~~~~~~~~~~~~~~~~~~~~~~~~~~~~~~~~~~~~~~~~~~~~~~~~}
\vspace*{13cm}

\noindent 
an extension rate twice as fast or twice as slow 
did not significantly modify 
the force-extension curves (data not shown).
In retrospect, this restricted set of unfolding pathways and their relative 
insensitivity to
 the precise values of external parameters also explain 
why we could simulate these force-induced unfolding pathways
starting from a given secondary structure, while predicting such
1540-nucleotide initial structure {\it a priori} is still beyond the 
current limitations of secondary structure prediction algorithms.

Despite clear similarities, the agreement between simulated and
 experimental
force-extension responses in Fig~8 is uneven.
In fact, variations between predicted (red) and measured (black) curves
might
reflect real differences between the probed structure and the actual 
native secondary structure inside the ribosome\cite{ramakrishnan,noller1}
used here as the intial structure in the simulations.
In particular, deviations at the beginning of the stretching plateau
might originate 
from  alternative base pair (re)folding 
of the 3' major domain ({\it III}) due to the absence of essential 
ribosomal proteins ({\it e.g.}, as s7\cite{nowotny})
and Mg$^{2+}$ ions.
Moreover, the relatively short time scales (few seconds) of these
stretching-induced unfolding/refolding experiments might not be
sufficiently long to let 16S rRNA find its lowest free energy structure
between successive pulls.

These results illustrate 
what should be expected, in general, when  RNA secondary structures
are probed by mechanical force. Strong helices resist until their breaking 
exposes weaker regions, which are unable to withstand such high 
forces. This 
leads to the unfolding of a significant domain with a concomitant force drop.
A fraction of the unpaired bases then typically reform different helices, 
which compensate, {\it in part}, for the sudden relaxation of the
mechanical tension. 
Yet, force-extension responses are {\it not completely} smoothed out, as 
initially suspected\cite{montanari,bundschuh,lubensky},  
by these local rearrangements.
Instead, they reveal the slow dynamics of large scale cooperative changes 
in complex RNA structures.
Tertiary interactions, likely marginal here due to the absence of 
Mg$^{2+}$ ions, are expected to strengthen the unfolding cooperativity 
between interacting domains and, concomitantly, increase the
reformation of non-native helices upon stretching.

Local rearrangements of RNA molecules, similar to those reported
here, likely occur {\it in vivo} as well, in particular, during 
translation when large domains of messenger RNAs become unfolded upstream 
of the ribosome. In fact, the influence of long-lived intermediate 
structures is likely ubiquitous to the RNA folding problem itself,
as slow structural rearrangements 
are known to occur in the context of both {\it in vitro} and 
{\it in vivo} RNA folding processes\cite{williamson1,woodson}. 
New experimental tools are needed to better
understand the strategies of RNA molecules in circumventing such kinetic 
traps (for instance through specific 
interactions with ions or proteins\cite{treiber}, 
through RNA chaperones\cite{woodson} or co-transcriptional encoded 
folding pathways\cite{isambert}).

By exploring RNA structure energy landscapes\cite{fontana},
micromanipulations combined with appropriate  stochastic 
simulations can help address such questions,  reflecting both 
structural and metastability features of single RNA molecules.

\vspace{0.6cm}
\noindent
{\bf Added note:}

Onoa {\it et al}\cite{onoa} have recently reported experimental results on 
the mechanical unfolding of the L-21 derivative of {\it Tetrahymena
  thermophila} ribozyme, a 390-nucleotide catalytic RNA. 
By contrast with the present study which strictly focuses on the RNA secondary 
structure level (no Mg$^{2+}$ added), 
Onoa {\it et al} primarily investigate the tertiary fold of this selfsplicing
ribozyme in  the  presence of Mg$^{2+}$.
A variety of hysteresis responses to the applied force is presented
for various parts of the molecule or in the presence of specific antisense 
oligos. A direct correlation between Mg$^{2+}$-dependent unfolding events 
and the opening of specific native helices is proposed.

\vspace{1cm}
\noindent
{\large \bf Materials and Methods}
{
\vspace{0.3cm}

\noindent
{\bf Sample displacement:} 
Sample displacement is driven and monitored by a nanometer resolution
piezoelectric stage with capacitive position sensor (P530-3, Physik
Instrument). The piezoelectric stage position is controlled and monitored by a
0-10V voltage. 
\vspace{0.3cm}

\noindent
{\bf Optical tweezer:} 
The optical tweezer consists of a Nd:Yag infra-red laser beam (TOPAZ,
SpectraPhysics) focussed inside the capillary by a 1.3 N.A. x100 objective
(Zeiss). The laser is always set at full power (2.5 W) and the
stiffness of the trap is controlled by the amplitude of an acoustic wave
generated by an 
acousto-optic modulator (A-A) placed right after the laser head. 
The experiments described
here were performed with a 50\% attenuation of the
laser intensity, which sets the optical trap stiffness around 
7$\times$10$^{-5}$~N/m.
The bead displacement from the laser beam focus point is measured as follows:
after passing through the sample, the bead diffused light is collected by a
0.6 N.A. x40 objective (Zeiss). The objective back focal plane is imaged by a
lense of 40 mm focal length onto a two-quadrant photodetector (S5980,
Hammamatsu). The whole experiment setup is mounted on an invar table so as to
minimize thermal position drift. The photodiode electric currents $I_A$ and
$I_B$ are converted into voltage and amplified by a home made amplifier.   
The voltage difference $V_A-V_B$ which is proportionnal to the distance of the
bead away from the trap center is further amplified and filtered at 300 Hz by
low noise amplifier (SR-50, Stanford Research Instrument). The total light
intensity that is collected by the x40 objective measured by the voltage sum
$V_A+V_B$ is also amplified. 
\vspace{0.3cm}

\noindent
{\bf Data acquisition:} 
The monitoring voltage coming out of the piezoelectric driver, the voltage
difference $V_A-V_B$ and the voltage sum $V_A+V_B$ coming out of the low noise
amplifier are each directed into a separate channel of an acquisition board
(ATMIO-16X, National Instrument). The driving voltage of the piezoelectric
stage is generated by the same board. The acquisition rate is 300 Hz which
sets the duration of the stretching/relaxing experiment around 10-20 seconds. 
\vspace{0.3cm}

\begin{figure}
\includegraphics{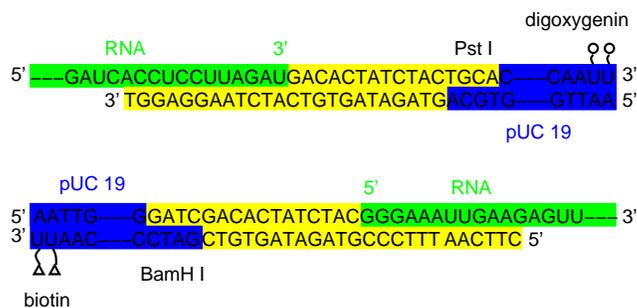}
\caption{\label{fig:wtgI} 
\small 
Detailed molecular junctions between RNA 3' end and digoxygenin labelled
pUC~19 (top) and between RNA 5' end and biotin labelled pUC~19
(bottom). Blue: pUC~19 DNA; Yellow: ssDNA oligos; Green: RNA insert.
}
\vspace{-.3cm}
\end{figure}

\noindent
{\bf Calibration:} 
The fourier power spectrum of a free bead inside the trap follows a lorentzian
law as expected for brownian fluctuations. Fitting this curve with two
parameters provides  both the trap stiffness and the voltage/distance
conversion factor. In the case of a pulling experiment, these two parameters
are used to convert the ratio $(V_A-V_B)/(V_A+V_B)$ directly into
piconewtons. The maximum force that can be measured with our setup is 60~pN. The bead position resolution inside the trap is $\pm$~5~nm which sets the
force resolution at $\pm$~0.4~pN.
The bead is captured at about 500-1000 nm from the capillary interior
surface. The pUC19 dimer contour length is 1742 nm (0.33 nm/pb). The geometry
imposes to displace the piezo stage by 48.8-56 nm and
985-1108 nm to completely unfolded the small RNA motifs and 16S rRNA, 
respectively.
\vspace{0.3cm}

\noindent
\mbox{\bf Molecule synthesis and functionalization:}~~RNA~molecu\-les were
synthesized by in vitro "run off" transcription of EcorV linearized DNA
plasmids. These plasmids were constructed by inserting DNA oligomers (IBA
GmBh) starting with a T7 promotor region inside the BamHI-PstI region of
pUC19. The RNA sequence was flanked by 12 nucleotides at both extremities to
allow for the ligation with the double-stranded DNA arm extensions. 
In the case of 16S rRNA, the gene was isolated by PCR from pKK3535 plasmid
(courtesy of K.~Lieberman and H.F.~Noller). It was cut by BstEII-BclI and 
then inserted in pUC19 together with oligomers carrying a T7 promotor 
with the DNA arm extensions and the complementary ends of the 16S
sequence. The reconstructed plasmids were produced in competent DH5 alpha
bacteria and were extracted and purified using Jetstar purification kit. They
were further sequenced. 
Due to the small length of RNA molecules, there were extended with digoxygenin
or biotin labelled dsDNA at, respectively, the 3' and 5' ends to enable
grafting between the capilary glass surface and the silica bead. In practice,
DNA oligomers (Fig~9) were first ligated to Pst~I restricted digoxygenin 
labelled pUC19 to yield a 12-nucleotide 3' extension complementary to the RNA
3' end. Then,
the DNA/RNA hybridization and ligation protocol was the following: RNA was
heated to 90$^{\circ}$C for 5 minutes then quenched on ice. It was incubated
with the former prepared pUC19 (molar ratio 100/1) at 70$^{\circ}$C for 20
minutes and then slowly cooled ($\ll$0.6$^{\circ}$C/min) to 16$^{\circ}$C. At
this temperature, T4 DNA ligase and buffer were added and the ligation
reaction was carried over 4 hours at 16$^{\circ}$C. The band corresponding to
the pUC19 molecular weight on a 0.7\% agarose gel was purified using Qiaquick
(Qiagen). The whole procedure was repeated with the RNA 5' end using a BamH~I
restricted biotin labelled pUC19 DNA  (Fig~9). The band corresponding to a
pUC19 dimer molecular weight on a 0.7\% agarose gel was purified using
Qiaquick (Qiagen). 
In the case of the 16S RNA, the functionalization protocol was slightly
modified. 
The oligomers were first hybridized with the RNA 3' end following the
heat-cooling protocol described above. The excess oligomers were washed away
by 2 consecutive centrifugations at 4000 $\times g$ and 16$^{\circ}$C using
GS-200 microspin column. The Pst~I restricted digoxygenin labelled pUC19 DNA,
T4 DNA ligase and buffer were added and the ligation reaction was carried over
4 hours at 16$^{\circ}$C. The same procedure was repeated with the 5' end and
the molecule was purified on an 0.7\% agarose gel by cutting the band
corresponding to a pUC dimer molecular weight. 
Prior to the experiment, the molecules are incubated with the streptavidin
coated beads (Bangs Laboratories) for 30 minutes. The solution is introduced
in the rectangular 
capillary by a peristatic pump which allows buffer circulation. All
experiments were performed at room temperature and in Tris 10~mM pH~7 NaCl 
250~mM buffer.

}

\vspace{0.6cm}
\noindent
{\large \bf Acknowledgements}

We thank K. Lieberman and H.F.~Noller for kindly 
providing us with the pKK3535 plasmid,   
D.~Evers and R.~Giegerich for the use of their ``RNAMovies'' 
software, and  L.~Bourdieu, C.~Ehresmann, S.~Lodmell, T.~Pan, M.~Poirier 
and  E.~Westhof  for discussions and suggestions. 
This work was supported in part by an ACI ``Jeunes Chercheurs'' grant from
Mi\-nist\`ere de la Recherche (France), an  NOI grant from the CNRS, and by
the ``Physique et Chimie du Vivant'' program of the CNRS.

\vspace{0.3cm}



\vfill
\eject

\end{document}